\documentclass[12pt]{article}

\input tcilatex
\QQQ{Language}{
American English
}

\begin{document}

\author{A.I.Volokitin$^{1,2}$ and B.N.J.Persson$^1$ \\
\\
$^1$Institut f\"ur Festk\"orperforschung, Forschungszentrum \\
J\"ulich, D-52425, Germany\\
$^2$Samara State Technical University, 443100 Samara,\\
Russia}
\title{Resonant photon tunneling enhancement of the van der Waals friction}
\maketitle

\begin{abstract}
We study the van der Waals friction between two flat metal surfaces in 
relative motion. For good conductors we find that normal relative motion gives
a much larger friction than for parallel
relative motion. The friction may increase  by many order of
magnitude when the surfaces are covered by adsorbates, or can support 
low-frequency surface plasmons. In this case
the friction is determined by resonant photon tunneling between adsorbate
vibrational modes, or surface plasmon modes.  
\end{abstract}

A great deal of attention has been devoted to non-contact friction between
nanostructures, including, for example, the frictional drag force between
two-dimensional quantum wells \cite{Gramila1,Gramila2,Sivan} , and the
friction force between an atomic force microscope tip and a substrate \cite
{Dorofeev,Gostmann,Stipe,Mamin,Hoffmann}. 

In non-contact friction the bodies are separated by a potential barrier
thick enough to prevent electrons or other particles with a finite rest mass
from tunneling across it, but allowing interaction via the long-range
electromagnetic field, which is always present in the gap between bodies.
The presence of inhomogeneous tip-sample electric fields is difficult to
avoid, even under the best experimental conditions \cite{Stipe}. For
example, even if both the tip and the sample were metallic single crystals,
the tip would still have corners present and more than one crystallographic
plane exposed. The presence of atomic steps, adsorbates, and other defects
will also contribute to the inhomogeneous electric field. The electric field
can be easily changed by applying a voltage between the tip and the
sample. 

The electromagnetic field can also be created by the fluctuating current density,
 due to
thermal and quantum fluctuations inside the solids. This fluctuating
electromagnetic field is always present close to the surface of any body,
and consist partly of traveling waves and partly of evanescent waves which
decay exponentially with the distance away from the surface of the body. The
fluctuating electromagnetic field originating from the fluctuating current 
density inside the bodies gives
rise to the well-known long-range attractive van der Waals interaction
between two bodies \cite{Lifshitz}. If the bodies are in relative motion, the
same fluctuating electromagnetic field will give rise to a friction which is
frequently named as the van der Waals friction.

Although the dissipation of energy connected with the non-contact friction
always is of electromagnetic  origin, the detailed mechanism is not
totally clear, since there are several different mechanisms of energy
dissipation connected with the electromagnetic interaction between bodies.
First, the electromagnetic field from one body will penetrate into the other
body, and induce an electric current. In this case friction is due to Ohmic
losses inside the bodies. Another contribution to friction from the
electromagnetic field is associated with the time-dependent stress acting
on the surface of the bodies. This stress can excite acoustic waves, or
induce time-dependent deformations which may result in a temperature
gradient. It can also induce motion of defects either in the bulk, or on the
surface of the bodies. The contribution to friction due to nonadiabatic heat
flow, or motion of defects, is usually denoted as internal friction.

It is very worthwhile to get a better understanding of different mechanisms
of non-contact friction because of it 
practical importance for
ultrasensitive force detection experiments. This is because the ability to
detect small forces is inextricably linked to friction via the
fluctuation-dissipation theorem. For example, the detection of single spins
by magnetic resonance force microscopy, which has been proposed for
three-dimensional atomic imaging \cite{Sidles} and quantum computation \cite
{Berman}, will require force fluctuations to be reduced to unprecedented
levels. In addition, the search for quantum gravitation effects at short
length scale \cite{Arkani} and future measurements of the dynamical Casimir
forces \cite{Mohideen} may eventually be limited by non-contact friction
effects.

Recently Gotsmann and
Fuchs  \cite{Gostmann} observed long-range non-contact friction between an aluminum
tip and
a gold (111) surface. The friction force $F$ acting on the tip is 
proportional to the velocity $v$, $F=\Gamma v$. For motion of the tip 
normal to the surface 
the  friction coefficient $\Gamma (d)=b\cdot d^{-3}$,
where $d$ is the tip-sample spacing and $b=(8.0_{-4.5}^{+5.5})\times
10^{-35}\mathrm{N\,s\,m}^2$  \cite{Gostmann}. Later Stipe \textit{et.al.}\cite{Stipe}
observed non-contact friction effect between a gold surface and a gold-coated
cantilever as a function of tip-sample spacing $d$, temperature $T$, and
bias voltage $V$. For vibration of the tip parallel to the surface they
found $\Gamma (d)=\alpha (T)(V^2+V_0^2)/d^n$, where $n=1.3\pm 0.2,$ and $%
V_0\sim 0.2\,\mathrm{V.}$ At 295\textrm{K, }for the spacing $d=$ 100\textrm{%
\AA\ }they found $\Gamma =1.5\times 10^{-13}\mathrm{\,kgs}^{-1},$ which is $%
\sim $500 times smaller that reported in Ref. \cite{Gostmann} at the same
distance using a parallel cantilever configuration.

In a recent Letter, Dorofeev \textit{et.al.} \cite{Dorofeev} claim that a
the non-contact friction effect observed in \cite{Dorofeev,Gostmann} is due to
Ohmic losses mediated by the fluctuating electromagnetic field . This result
is controversial, however, since the van der Waals friction has
been shown \cite{Volokitin1,Persson and Volokitin,Volokitin2,Volokitin3} to
be many orders of magnitude smaller than the friction  observed by Dorofeev \textit{%
et.al.} Presently, the origin of the difference in magnitude and distance
dependence of the long-range non-contact friction effect observed in \cite
{Gostmann} and \cite{Stipe} is not well understood.

In order to improve the basic understanding of  non-contact friction,
in this Letter we present new results for van der Waals friction.  
In \cite{Persson and Volokitin} we developed a theory of van der Waals friction
for surfaces in parallel relative motion.
Here we generalize the theory to include also the case when the 
surfaces are in 
normal relative motion, and we show that there is drastic difference between
these two cases. Thus,  for normal relative motion of clean good conductor surfaces,
the friction is many orders of magnitude larger than for parallel relative motion, 
but  still  smaller than observed  experimentally.
Another enhancement mechanism of the non-contact friction can be connected
with resonant photon tunneling between  states localized on the different
surfaces. Recently it was discovered that resonant photon tunneling between
surface plasmon modes give rise to extraordinary enhancement of the optical
transmission through sub-wavelength hole arrays \cite{Ebbesen}. The same
surface modes enhancement  can be expected  for van der Waals friction if the
frequency of these modes is sufficiently low to be excited by thermal radiation.
At room temperature only the modes with frequencies below 
$\sim 10^{13}s^{-1}$ can be excited.
For normal metals surface plasmons have much too high frequencies; at 
thermal frequencies 
the dielectric function of normal metals becomes nearly
purely imaginary, which exclude surface plasmon enhancement of the van der Waals
friction for good conductors. However surface plasmons 
for semiconductors are characterized by much
smaller frequencies and damping constants, and they can give an important contribution
to van der Waals friction. 
Other  surface modes which can be excited 
by thermal radiation are
adsorbate vibrational modes. Especially for parallel vibrations these modes
may have very low frequencies.

Recently \cite{Volokitin1} we developed a theory of the van der Waals friction
between two-semiinfinite bodies, moving parallel to each other.
We have  generalized this theory  
to two-semiinfinite bodies, moving normal to each other. The frictional  stress,
$\sigma$, is proportional to the velocity $v$, $\sigma=\gamma v$.
 
For the separation  $d<c\hbar/k_BT$ 
we get coefficient of friction $\gamma_{\perp}$ \cite{Volokitin4}:
\begin{eqnarray}
\gamma _{\perp }=\frac \hbar {\pi ^2}\int_0^\infty d\omega \int dqq^3\left( -
\frac{\partial n(\omega )}{\partial \omega }\right) e^{-2qd} \nonumber \\
\times \left\{((\mathrm{Im}R_{1p}+e^{-2qd}\left| R_{1p}
\right| ^2\mathrm{Im}R_{2p})(\mathrm{Im}R_{2p}+e^{-2qd}\left|
R_{2p}\right| ^2\mathrm{Im}R_{1p} ) \right. \nonumber \\
\left.+ e^{-2qd}(\mathrm{Im}(R_{1p}R_{2p}))^2)\frac {1}{\left|
1-e^{-2qd}R_{1p}R_{2p}\right| ^4}
+[p\rightarrow s] \right\}   \label{one}
\end{eqnarray}
where the Bose-Einstein factor
\begin{equation}
n(\omega)=\frac {1}{e^{\hbar \omega/k_BT}-1}
\end{equation}
The symbol $[p\rightarrow s]$ in (\ref{one}) denotes the term which is
obtained from the first one by replacement of the reflection factor $%
R_p(\omega )$ for $p-$ polarized waves by the reflection factors $%
R_s(\omega )$ for $s-$ polarized waves. There is a principal difference
between the friction coefficient for normal and parallel relative motion, 
related to the denominator in (\ref{one}).  
The resonant condition corresponds to the case when the denominator of the integrand
in (\ref{one}), which is due to multiple scattering of evanescent electromagnetic
waves from opposite surfaces, is small. For two identical surfaces and 
$R_{i}<<1\le R_r$,
where $R_i$ and $R_r$ are the imaginary and real part, respectively, this
corresponds to the resonant condition $R_r^{2}\rm{exp}(-2qd)\approx 1$. At 
resonance the integrand in (\ref{one}) has a large factor $\sim 1/R_i^2$,
in  sharp contrast to the case of parallel relative motion, where  there is no such
enhancement factor. The resonance condition can be fullfiled even for the case 
when $\rm{exp}(-2qd)<<1$ because for evanescent electromagnetic waves there is no
restriction on the magnitude of real part or the modulus of $R$. 
This open up the possibility of 
resonant denominators for $R_r^2>>1$.
 
The reflection factor $R_{p}$, which
take into account the contribution from an adsorbate layer, is given by \cite{Langreth}:
\begin{equation}
R_{p}=\frac {q-s/\epsilon+4\pi n_aq[s\alpha_{\parallel}/\epsilon+q \alpha_{\perp}]}
{q+s/\epsilon+4\pi n_aq[s\alpha_{\parallel}/\epsilon-q \alpha_{\perp}]}, \label{two}
\end{equation} 
where 
\begin{equation}
s=\sqrt{q^2-\left(\frac{\omega}{c}\right)^2\epsilon},   \\
\end{equation}
and where $\alpha_{\parallel}$ and $\alpha_{\perp}$ are the  
polarizabilities of adsorbates
in a direction parallel and normal to the surface,  respectively. $\epsilon$
is the bulk  dielectric function and $n_a$ is
the concentration of adsorbates. For clean surfaces $n_a=0$, and in this
case formula (\ref{two}) reduces to the well-known Fresnel formula. 

Let us
first consider two identical metal described by the dielectric function
\begin{equation}
\epsilon=1-\frac {\omega_p^2}{\omega (\omega+i\tau^{-1})},
\end{equation}
where $\tau$ is the relaxation time and $\omega_p$ the plasma frequency.
For good
conductors at thermal frequencies $R_{pi}<<1$ and $R_{pr}\approx1$. Thus an
enhancement in friction is possible only for very small $q<<1/d$ .
 Analysis show that integral (\ref{one}) has a
$1/q^3$-singularity  and the main contribution to the integral (\ref{one})
comes from vicinity of this singularity. 
For two bodies,
moving parallel to each other, the integral has only a logarithmic singularity 
for small $q$,  and the main contribution comes from the non-resonant 
region with $q\sim 1/d$. Thus, 
for clean surfaces of good conductors, for normal relative motion 
the van der Waals friction will be on many order 
of magnitude larger than for  parallel relative motion.
Fig.1 illustrates this situation for two copper surfaces and $T=273K$. However
the van der Waals friction in this case is too small in 
comparison with experimental data. Thus,  for 
$d=1nm$ the friction $\gamma_{\perp theor} 
\sim 10^{-4}kgs^{-1}m^{-2}$. For an atomic force microscope one
 tip 
estimate $\Gamma \approx \gamma S$, where $S \approx Rd$ 
is an effective surface area of the tip with a radius of curvature $R$. For the tip
with $R \sim 1\mu m$ the friction coefficient 
$\Gamma_{theor} \sim 10^{-19}kgs^{-1}$, while from the experimental data at the same 
distance one can deduce
the friction coefficient $\Gamma_{exp} \sim 10^{-12}kgs^{-1}$ \cite{Stipe}.

Resonant photon tunneling enhancement of the van der
Waals friction is possible for two semiconductor surfaces which can support
low-frequency surface plasmon modes. As an example we consider 
two clean surfaces of silicon carbide (SiC). The
optical properties of this material can be described using an oscillator
model \cite{Palik}
\begin{equation}
\epsilon(\omega)=\epsilon_{\infty}\left(1+\frac{\omega_L^2 - \omega_T^2}
{\omega_T^2 - \omega^2 -i\Gamma \omega}\right)
\label{three}
\end{equation}
with $\epsilon_{\infty}=6.7, \,\omega_L=1.8\cdot10^{14}s^{-1},\,
\omega_T=1.49\cdot10^{14}s^{-1},\,$ and $\Gamma=8.9\cdot10^{11}s^{-1}.$ The
frequency of surface plasmons is determined by condition
$\epsilon_r(\omega_p)=
-1$ and from (\ref{three}) we get $\omega_p=1.78\cdot10^{14}s^{-1}$. In Fig.2
we plot the friction coefficient $\gamma(d)$: note that the
friction between the two semiconductor surfaces is  several order of magnitude
larger than between two clean  good conductor surfaces.

Another enhancement mechanism is connected with resonant photon tunneling
between adsorbate vibrational modes localized on different surfaces. In this 
case the real part of the reflection factor $R_{p}$ can be much larger than unity, 
and the resonant condition $\rm{exp}(-2qd)R_p^2 \sim 1$ can be fulfilled even for 
large $q$, giving rise to large enhancement of friction. 
As an example, let us consider ions with charge $e^*$ adsorbed on metal surfaces. 
The polarizability for ion vibration normal to the surface is given by
\begin{equation}
\alpha_{\perp}=\frac {e^{*2}}{M(\omega_{\perp}^2-\omega ^2 -i\omega \eta_{\perp})},
\end{equation}  
where $\omega_{\perp}$ is the frequency of the normal adsorbate  vibration, 
and $\eta_{\perp}$ is the damping constant.      
In Eq. (\ref{two}) the contribution  from parallel vibrations 
is reduced by the small factor $1/\epsilon$. However, the contribution
of parallel vibrations to the van der Waals friction
 can nevertheless   be important 
due to the indirect interaction of parallel adsorbate vibration with the electric field,
via the metal conduction electron \cite{Persson and Volokitin2}. Thus, the small 
parallel component
of the electric field will induce a strong electric current in the metal. 
The drag force between
the electron flow and adsorbates can induce adsorbate vibrations  parallel to the
surface. This gives the polarizability:
\begin{equation}
\alpha_{\parallel}=\frac {\epsilon -1}{n}\frac {e^{*}}{e}\frac {\omega \eta_{\parallel}}
{(\omega^2_{\parallel}-\omega^2 -i\omega \eta_{\parallel})}
\end{equation}
where $n$ is the conduction electron concentration.
As an illustration, in Fig.3 we show coefficient of friction for the two  
Cu(001) surfaces covered by a low concentration of potassium atoms
( $n_a=10^{18}m^{-2}$) 
. In  the $q-$ 
integral in Eq.(\ref{one}) we used the cut off  $q_c \sim \pi/a$ (where $a\approx1nm$ 
is the inter-adsorbate
distance) because our microscopic approach is applicable only when the wave length
of the electromagnetic field is larger than double average distance between the
adsorbates.
 In comparison, the friction  
between two clean surface at separation $d=1nm$ is seven order of 
magnitude smaller. 
 At $d=1nm$
the friction coefficient $\Gamma$ for an atomic force microscope tip 
with $R \sim 1\mu m$
is $\sim 10^{-12}kgs^{-1}$ ($\gamma \sim 10^{3}kgs^{-1}m^{-2}$, see
Fig.2);  
this is of the same order of magnitude 
as the  observed friction \cite{Stipe}.

In this letter, we have shown that the  van der Waals friction can be enhanced 
by several orders of magnitude when the material involved support low-frequency
adsorbate vibrational modes or surface plasmon modes. For clean surfaces of 
good conductors 
friction is  several order of magnitude larger for normal relative motion 
as compared to parallel relative
motion. These results should have broad  application in non-contact friction
microscopy, and  in the design of new tools for studying adsorbate vibrational dynamics.

\vskip 0.5cm \textbf{Acknowledgment }

A.I.V acknowledges financial support from DFG. B.N.J.P. acknowledges 
support from the European Union Smart Quasicrystals project. 
\vskip 1cm

\vskip 0.5cm

FIGURE CAPTIONS

Fig. 1. The friction coefficient for two (clean) surfaces 
in (a) normal and (b) parallel relative motion, as a function of
separation $d$. $T=273$K and with parameters chosen to correspond to copper (
$ \tau ^{-1}=2.5\cdot 10^{13}s^{-1},\,\omega _p=1.6\cdot 10^{16}s^{-1}$). 
 (The log-function is with
basis 10)

Fig.2. The friction coefficient for two  clean semiconductor surfaces
in (a) normal and (b) parallel relative motion, as a function of the
separation $d$. $T=300$K and  with parameters chosen to correspond to
a surfaces of silicon carbide (SiC) (see text for explanation)
(The log-function is with basis 10)

Fig. 3. The friction coefficient for two  surface covered by adsorbates
 in (a) normal and (b) parallel relative motion, as a function of the
separation $d$. $T=273$K and  with parameters chosen to correspond to K/Cu(001)
\cite{Senet} ($ \omega_{\perp}=1.9\cdot 10^{13}s^{-1}, \omega_{\parallel}=
4.5\cdot 10^{12}s^{-1}, \eta_{\parallel}=2.8\cdot 10^{10}s^{-1},
 \eta_{\perp}=1.6\cdot 10^{12}s^{-1}, e^{*}=0.88e$)
(The log-function is with basis 10)

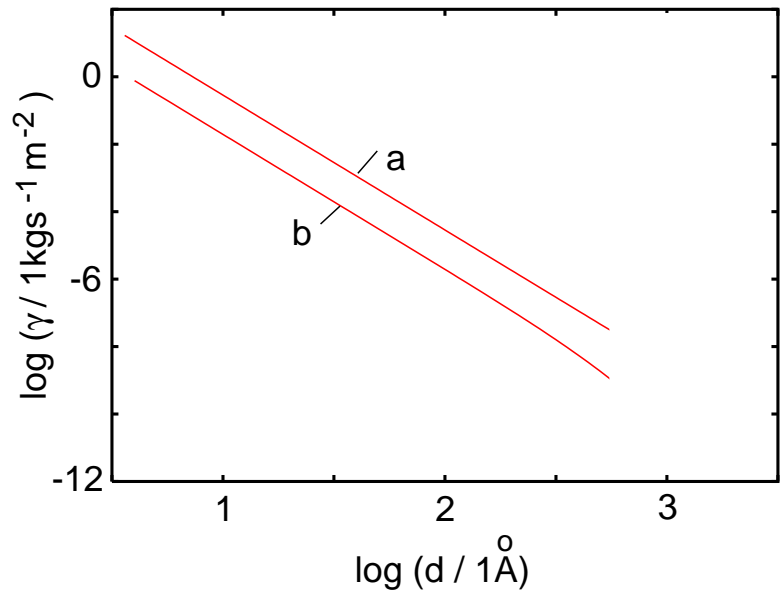
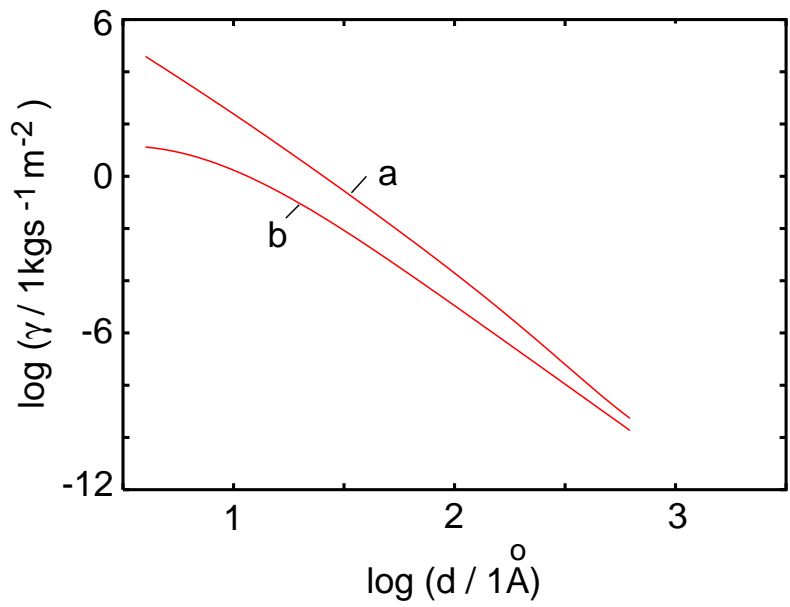
\end{document}